\documentclass[journal]{IEEEtran}  

\IEEEoverridecommandlockouts                              

\usepackage{graphics} 
\usepackage{epsfig} 
\usepackage{times} 
\usepackage{amsmath} 
\usepackage{amssymb}  
\usepackage{enumitem}
\usepackage{xcolor}
\usepackage{bbm}
\usepackage{bm}
\usepackage{cite}
\usepackage{tikz}
\usetikzlibrary{mindmap,shadows,arrows,decorations.pathmorphing,backgrounds,positioning,fit,petri}
\usepackage[hyphens]{url}

\newtheorem{theorem}{Theorem}

\newtheorem{remark}[theorem]{Remark}

\title{\LARGE \bf
Model Predictive Control-Based Battery Scheduling and Incentives to Manipulate Demand Response Baselines 
}

\author{Douglas Ellman and Yuanzhang Xiao
\thanks{The authors are with Hawaii Advanced Wireless Technologies Institute and Department of Electrical Engineering, University of Hawaii at Manoa, Honolulu, HI 96822, USA. Email:\{dellman, yxiao8\}@hawaii.edu.}%
}

\begin{document}

\maketitle
\thispagestyle{empty}
\pagestyle{empty}

\begin{abstract}
We study operations of a battery energy storage system under a baseline-based demand response (DR) program with an uncertain schedule of DR events. Baseline-based DR programs may provide undesired incentives to inflate baseline consumption in non-event days, in order to increase ``apparent'' DR reduction in event days and secure higher DR payments. Our goal is to identify and quantify such incentives. 
To understand customer decisions, we formulate the problem of determining hourly battery charging and discharge schedules to minimize expected net costs, defined as energy purchase costs minus energy export rebates and DR payments, over a sufficiently long time horizon (e.g., a year).
The complexity of this stochastic optimization problem grows exponentially with the time horizon considered. To obtain computationally tractable solutions, we propose using multistage model predictive control with scenario sampling. Numerical results indicate that our solutions are near optimal (e.g., within 3\% from the optimum in the test cases).
Finally, we apply our solutions to study an example residential customer with solar photovoltaic and battery systems participating in a typical existing baseline-based DR program. Results reveal that over 66\% of the average apparent load reduction during DR events could result from inflation of baseline consumption during non-event days.
\end{abstract}


\section{Introduction} \label{Introduction}
An important mechanism for improving reliability and reducing costs of the power grid is demand response (DR). Among the various DR mechanisms that have been proposed or implemented \cite{Vardakas2015}\cite{SmartElectricPoweralliiance2018}, we can categorize most of them as price-based DR and incentive-based DR. Price-based DR uses time-varying electricity prices to encourage customers to reduce consumption when prices are high. Alternatively, incentive-based DR pays customers rebates based on participation or performance in the DR program. 

This paper focuses on baseline-based DR 
\cite{Dobakhshari2018,Muthirayan2019,muthirayan2017mechanism,Vuelvas2018,Vuelvas2018a,Ellman2019,Bruno2018,Garifi2018,Castelo-Becerra2017}, an important and widely-used class of incentive-based DR. Baseline-based DR is prevalent in practice, because it can be an opt-in program without changing customer electricity rates, thus making it easier to achieve regulatory approval. Baseline-based DR pays customers based on the difference between energy consumption during DR events and baseline consumption (e.g., average consumption during previous non-event days).

While baseline-based DR gives customers incentives to reduce load during DR events, it may also create undesired incentives for customers to manipulate baselines. Randomized control trials \cite{Wolak2006} and legal settlements \cite{JimPierobon2013} have provided real-world evidence that customers may manipulate baselines. 
In the past, manipulating DR baselines may have required substantial customer effort or diminished customer comfort (e.g., manually changing air conditioner settings). But today, battery systems with automated controllers can optimize customer load profiles to maximize DR revenues with minimal customer effort and discomfort. Thus, customers with batteries may automatically use sophisticated strategies to exploit incentives to manipulate baseline load, perhaps without being aware that they are doing so.
These behaviors may cause the utility to make significant errors in projecting demand forecasts, required DR quantities, and DR costs. Therefore, it is important to identify such behaviors and quantify their impacts.

To identify and quantify incentives to manipulate baselines, we need to understand how the customer schedules hourly battery charging and discharging when participating in baseline-based DR programs over a sufficiently long time horizon (e.g., a year), facing the uncertainty in when will the DR events occur. We model the customer's decisions as solutions to a stochastic optimization problem of minimizing the expected net costs, defined as the energy purchase costs minus energy export payments and DR payments. However, the stochastic optimization problem is intractable to solve, because its complexity grows exponentially with the time horizon. We propose a multistage model predictive control (MPC) approach, which determines the battery schedule of each day by solving a stochastic program over a much shorter receding horizon (e.g., a month). Through simulations, we demonstrate that our proposed solutions are near optimal (e.g., within 3\% from the optimum in the test cases). Then we apply our solutions to study an example residential customer with solar photovoltaic and battery systems participating in a typical existing baseline-based DR program. Results indicate that over 66\% of the average apparent load reduction during DR events could result from inflation of baseline consumption during non-event days. 

We summarize our major contributions as follows:
\begin{itemize}
   \item our work is the first to study incentives to manipulate DR baselines by using a battery;
   \item our work introduces a stochastic optimization problem formulation, which is general enough to model a variety of baseline-based DR mechanisms over a long time horizon with uncertain DR event schedules;
   \item our work proposes a computationally-tractable and near-optimal solution to the problem formulated, which allows us to identify and quantify customer incentives.
\end{itemize}

The remainder of this paper is organized as follows: Section II discusses related work, Section III describes the problem, Section IV describes our solution, Section V describes a numerical case study for an example Honolulu residential customer, and Section VI summarizes conclusions.  

\section{Related Work} \label{Related}
There is a huge literature on optimal battery dispatch under price-based DR (see some representative works in 
\cite{Vedullapalli2019, Vrettos2013, Keerthisinghe2019, Jin2017, Xu2014}).
Some of these works use solution concepts similar to ours, namely stochastic programming and model-predictive control. However, since baselines are not part of price-based DR, these works do not study incentives to game baselines.

Some works study optimal battery dispatch under baseline-based DR, but use pre-established baselines as inputs to their models \cite{Bruno2018,Garifi2018,Castelo-Becerra2017}. These works essentially assume that the customer does not have the ability to manipulate the baseline, and therefore do not study incentives to game baselines. 

Many of the works that study gaming behavior focus on mechanism design of the DR program \cite{Dobakhshari2018,Muthirayan2019,muthirayan2017mechanism,Vuelvas2018}. In \cite{Muthirayan2019,muthirayan2017mechanism,Vuelvas2018}, customers self-report baselines, and truthful reporting is encouraged by excluding a subset of customers from each event, and imposing penalties for deviations from reported baselines for those non-participating customers. In \cite{Dobakhshari2018}, there is a profit-sharing mechanism designed to discourage manipulating baselines for a DR program in which the baseline is the load at the start of the DR event. These works propose new DR mechanisms and do not model battery operations. In contrast, our work studies optimal battery operations under more commonly used existing DR mechanisms.

The most related works are \cite{Vuelvas2018a}\cite{Ellman2019}, which study gaming incentives in baseline-based DR. The work in \cite{Vuelvas2018a} only includes results for a two-stage model, where the second stage is the event day with certainty.
In comparison, our work considers uncertain event schedules over many days. 
Our prior work \cite{Ellman2019} studies a multi-day DR season with uncertain DR event schedules. It uses dynamic programming with backwards induction to obtain optimal energy consumption during DR windows of event and non-event days, where there is a discrete set of daily feasible consumption levels and the costs of realizing those consumption levels are known and independent of actions on other days. In contrast, our current work considers a battery system, where the feasible hourly charging and discharging quantities and associated opportunity costs depend on actions taken at other times, and the action space is continuous. 
We use model predictive control (MPC) to get a near-optimal battery dispatch, because the complexity of the model in \cite{Ellman2019} grows polynomially with the granularity of the discretization of inherently continuous action and state spaces, while our MPC approach allows continuous states and actions for free. 
Additionally, this work adds representation of DR mechanisms where payments depend on average power (kW) reduction over multiple events (DR capacity payments). DR capacity payments could not be easily considered by the model in \cite{Ellman2019} because it would require significantly enlarging the state space to include information about prior events. 


\section{System Model and Problem Formulation} \label{MODEL}
\subsection{Model Setup}
We consider a customer with electric loads, a solar photovoltaic system, and a battery energy storage system. The customer participates in a baseline-based demand response program and is subject to a certain electricity tariff. We look at a finite horizon of $T$ days, where each day is indexed by $t \in \{1,\ldots,T\}$. Due to the structure of the DR payments, these $T$ days are divided into $I$ intervals (e.g., 12 months). The time-granularity of the battery charging and discharging decisions considered here is one hour, where each hour is indexed by $h \in \{1,\ldots,H=24 \cdot T \}$. As will be useful later, we write $t_i$ as the final day of each interval $i \in \{1,\ldots,I\}$, and $h_t$ as the final hour of each day $t \in \{1,\ldots,T\}$. Our objective is to determine the optimal schedule of battery charging and discharging in each hour to minimize the expected total net cost for the customer over the $T$ days.

The demand response program aims to incentivize the customer to reduce electricity consumption during a pre-specified window of hours (e.g., 5 p.m -- 9 p.m.) on demand response \textit{event days}. Prior to any day, there is uncertainty on whether that day is an event day. This uncertainty can be represented by a Bernoulli random variable $\omega_t$ with parameter $P_t$, where $\omega_t=1$ indicates that day $t$ is an event day, and the probability that day $t$ is an event day is $P_t$. We assume that the random variables $\omega_1,\ldots,\omega_T$ are independent. 
At the beginning of each day, the customer is notified whether the current day is an event day. 

In each hour $h$, the customer has default electricity demand $d_h$ and solar energy production $\rho_h$. The customer can charge the battery by an amount of $b^+_h \geq 0$ or discharge the battery by an amount of $b^-_h \geq 0$,
resulting in an hourly net load 
$$
l_h = d_h + b^+_h - \rho_h - b^-_h.
$$ 
Positive values of $l_h$ indicate that the customer purchases energy from the grid during hour $h$, and negative values indicate that the customer exports energy to the grid. The values $d_h$, $\rho_h$, $b_h$, and $l_h$ are in the unit of kWh.

The battery scheduling variables $b^+_h$ and $b^-_h$ are constrained by the physical characteristics of a battery system. A power capacity constraint requires the total amount of energy charged and discharged to be no larger than the rated power capacity $P$ of the battery system in all hours, namely
\begin{eqnarray} \label{eqn:power_capacity_constraint}
b^+_h + b^-_h \leq P,~\forall h \in \{1,\ldots,H \}.
\end{eqnarray}
We denote the level of stored energy at each hour $h$ by $e_h$. The dynamics of the stored energy can be written as
\begin{eqnarray} \label{eqn:battery_dynamics}
e_h = e_{h-1} + b^+_h \cdot \eta^+ - b^-_h / \eta^-,~\forall h \in \{2,\ldots,H \},
\end{eqnarray}
where $\eta^+ \in (0,1)$ is the battery charging efficiency and $\eta^- \in (0,1)$ is the battery discharging efficiency.
Finally, an energy capacity constraint requires that the level of stored energy does not exceed the rated energy capacity $E$ of the battery system in all hours, namely
\begin{eqnarray} \label{eqn:energy_capacity_constraint}
0 \leq e_h \leq E,~\forall h \in \{1,\ldots,H \}.
\end{eqnarray}

We define $\bm{b}_t^+ = (b^+_{h_{t-1}+1},\ldots,b^+_{h_t})$ as the vector of battery charging schedules in day $t$, and $\bm{b}_t^- = (b^-_{h_{t-1}+1},\ldots,b^-_{h_t})$ as the vector of battery discharging schedules in day $t$.

\subsection{Costs and Payments}
The customer's economic incentives consist of costs and payments. The cost is the customer's energy purchase cost.
For each hour when the net load is positive (i.e., $l_h > 0$), the customer incurs an energy purchase cost $c_h$, calculated as
\begin{eqnarray} \label{eqn:ckwh}
c_h\left(b^+_h, b^-_h\right) = r^{c}_h \cdot l_h = r^{c}_h \cdot \left( d_h + b^+_h - \rho_h - b^-_h \right),
\end{eqnarray}
where $r^{c}_h$ is the electricity purchase rate per kWh for hour $h$. 

The payments come from energy export and the DR program. For each hour when the net load is negative (i.e., $l_h < 0$), the customer exports energy to the grid and receives an energy export payment $p^{e}_h$, calculated as
\begin{eqnarray} \label{eqn:pkwh}
p^{e}_h\left(b^+_h, b^-_h\right) = r^{e}_h \cdot (-l_h) = r^{e}_h \cdot \left( - d_h - b^+_h + \rho_h + b^-_h \right),
\end{eqnarray}
where $r^{e}_h$ is the electricity export rate per kWh for hour $h$.

Additionally, the customer may receive DR energy payments or DR capacity payments or both via the demand response program. Both types of DR payments are based on the quantity of demand response energy reduction, which is the difference between the true energy consumption and the {\it baseline} energy consumption during the DR window. We write
$\mathcal{H}_t$ as the set of hours in the DR window of day $t$. Then the true energy consumption during the DR window of day $t$, denoted by $s_t$, is calculated as
\begin{eqnarray}\label{eqn:consumption}
s_t = \sum_{h \in \mathcal{H}_t} l_h.
\end{eqnarray}
At an event day $t$, the baseline consumption is calculated based on the consumption during the DR windows of a number of previous non-event days. We write $\mathcal{T}_t^B$ as the set of non-event days that affect the baseline in the event day $t$. Then the baseline energy consumption in an event day $t$, denoted by $\bar{s}_t^B$, is defined as
\begin{eqnarray}\label{eqn:baseline_consumption}
\bar{s}_t^B = f( \{s_\tau\}_{\tau \in \mathcal{T}_t^B}  ).
\end{eqnarray}
Usually, the baseline is simply the average consumption during the DR windows of relevant non-event days (e.g. \cite{HawaiianElectric2018}), namely
\begin{eqnarray*}
\bar{s}_t^B = \left( \textstyle\sum_{\tau \in \mathcal{T}_t^B} s_\tau \right) / \left|\mathcal{T}_t^B\right|.
\end{eqnarray*}
Given the true and baseline consumption, the DR energy reduction in event day $t$, denoted by $\Delta_t$, can be calculated as
\begin{eqnarray}\label{eqn:reduction}
\Delta_t = \bar{s}_t^B - s_t.
\end{eqnarray}
Note that we allow the reduction to be negative, so the customer could potentially pay a penalty for increasing the demand, as in \cite{HawaiianElectric2018}. Other programs do not penalize the customer for increasing the demand, in which case we will have $\Delta_t = \max\left\{0, \bar{s}_t^B - s_t \right\}$.

As in \cite{ConEdison2019}, the DR energy payment for each event day $t$, denoted by $p^{DRe}_t$, is calculated by 
\begin{eqnarray}\label{eqn:pdrkwh}
p^{DRe}_t\left(\bm{b}^+_{1:t},\bm{b}^-_{1:t};\bm{\omega}_{1:t}\right) = r^{DRe}_t \cdot \Delta_t,
\end{eqnarray}
where $r^{DRe}_t$ is the DR payment rate per kWh for day $t$, $\bm{b}^+_{1:t}=(\bm{b}^+_1,\ldots,\bm{b}^+_{t})$ and $\bm{b}^-_{1:t}=(\bm{b}^-_1,\ldots,\bm{b}^-_{t})$ are the battery charging and discharging schedules from day $1$ to day $t$, respectively, and $\bm{\omega}_{1:t}=(\omega_1,\ldots,\omega_t)$ is the sequence of event indicators up to day $t$. The customer receives no DR energy payment on non-event days. 

\begin{remark}[Inter-Temporal Dependence]\label{remark:dependence}
The DR energy payment for an event day $t$ could depend on the battery scheduling decisions in day $1$, if for example, days $2$ to $t-1$ are all event days. We write $p^{DRe}_t\left(\bm{b}^+_{1:t},\bm{b}^-_{1:t};\bm{\omega}_{1:t}\right)$ to explicitly indicate the inter-temporal dependence of the payment on the scheduling decisions. Strictly speaking, the DR energy payment depends only on the hours {\it in the DR windows} of previous days. But we use battery scheduling decisions of all previous hours $\bm{b}^+_{1:t}$ and $\bm{b}^-_{1:t}$ to simplify notations.
\end{remark}

\begin{remark}[Randomness]\label{remark:randomness}
In \eqref{eqn:pdrkwh}, we also make it clear that the DR payment $p^{DRe}_t$ depends on the realization $\bm{\omega}_{1:t}$ of the event days from day $1$ to day $t$, and therefore is random. This is because the set $\mathcal{T}_t^B$ of non-event days that affect the baseline in day $t$ depends on $\bm{\omega}_{1:t}$ and is random.
As a consequence, all the quantities related to DR events, namely $s_t$, $\bar{s}_t^B$, $\Delta_t$, and $p_t^{DRe}$ depend on $\bm{\omega}_{1:t}$ and are random.
\end{remark}

Another type of DR payment is the DR capacity payment, which is based on the average energy reduction {\it per hour} during an interval of days (e.g., a month). We write the set of event days in interval $i$ as $\mathcal{T}_i^E$. Then the average energy reduction per hour during interval $i$, denoted by $\bar{\Delta}_i$, is calculated as
\begin{eqnarray}\label{eqn:average_reduction_per_interval}
\bar{\Delta}_i = \frac{\sum_{t \in \mathcal{T}_i^E} \Delta_t}{\sum_{t \in \mathcal{T}_i^E} \left|\mathcal{H}_t\right|}.
\end{eqnarray}
As in \cite{ConEdison2019}, the DR capacity payment during interval $i$ is then 
\begin{eqnarray} \label{eqn:pdrkw}
p^{DRc}_i\left(\bm{b}^+_{1:{t_i}},\bm{b}^-_{1:{t_i}};\bm{\omega}_{1:t_i}\right) = r^{DRc}_i \cdot \bar{\Delta}_i,
\end{eqnarray}
where $r^{DRc}_i$ is the DR capacity payment rate for interval $i$, and $\bm{b}^+_{1:{t_i}}$ and $\bm{b}^-_{1:{t_i}}$ are the battery charging and discharging schedules from day $1$ to the final day $t_i$ of interval $i$, and $\bm{\omega}_{1:t_i}=(\omega_1,\ldots,\omega_{t_i})$ is the sequence of event indicators up to the final day $t_i$ of interval $i$. Again in \eqref{eqn:pdrkw}, we make it clear that the DR capacity payment $p^{DRc}_i$ depends on prior scheduling decisions and is random, as explained in Remarks~\ref{remark:dependence}--\ref{remark:randomness}. If there is no DR event during interval $i$, we set $\bar{\Delta}_i = 0$. Note that some DR programs do give DR capacity payments (e.g., the prior interval's amount) in the case of no event.


\subsection{Problem Formulation}

We aim to minimize the customer's expected total net cost over a finite time horizon of $I$ intervals (e.g., 12 months), which consist of $T$ days (e.g., 365 days), or equivalently, $H=24\cdot T$ hours (e.g., 8760 hours). The decision variables are battery scheduling variables $\bm{b}_{1:T}^+$ and $\bm{b}_{1:T}^-$ during this time horizon. 

As shown in \eqref{eqn:pdrkwh} and \eqref{eqn:pdrkw}, the DR payments depend on the realization of random event days. Therefore, the battery schedules $\bm{b}^+_{t}$ and $\bm{b}^-_{t}$ in day $t$ should also depend on the previous and current realizations $\bm{\omega}_{1:t}$ of event days (but not future realizations due to causality). To make this dependence explicit, we write the battery schedules as functions of the realizations, namely $\bm{b}^+_{t}(\bm{\omega}_{1:t})$ and $\bm{b}^-_{t}(\bm{\omega}_{1:t})$.

Given realizations of all events $\bm{\omega}_{1:T}$, the total net cost is
\begin{eqnarray} \label{eqn:net_cost}
& & C\left( \left\{\bm{b}^+_{t}(\bm{\omega}_{1:t}), \bm{b}^-_{t}(\bm{\omega}_{1:t}) \right\}_{t=1}^T ; \bm{\omega}_{1:T} \right) \\
&=& \sum_{h=1}^{H} (c_h - p^e_h) - 
    \sum_{i=1}^I p^{DRc}_i\left(\bm{b}^+_{1:{t_i}}(\bm{\omega}_{1:t}),\bm{b}^-_{1:{t_i}}(\bm{\omega}_{1:t});\bm{\omega}_{1:t_i}\right) \nonumber \\
& & - \sum_{i=1}^I \sum_{t \in \mathcal{T}_i^E} p^{DRe}_t\left(\bm{b}^+_{1:t}(\bm{\omega}_{1:t}),\bm{b}^-_{1:t}(\bm{\omega}_{1:t});\bm{\omega}_{1:t}\right). \nonumber
\end{eqnarray}
Our goal is to minimize the expectation of the above cost:
\begin{eqnarray} \label{eqn:optimization_problem}
&\text{minimize}  &
\mathbb{E}_{\bm{\omega}_{2:T}} \left\{
C\left( \left\{\bm{b}^+_{t}(\bm{\omega}_{1:t}), \bm{b}^-_{t}(\bm{\omega}_{1:t}) \right\}_{t=1}^T ; \bm{\omega}_{1:T} \right) 
\right\} \nonumber \\
&\text{subject to:}& \text{under each}~\bm{\omega}_{2:T} \in \{0,1\}^{T-1}, \nonumber \\
&                  & \text{for}~t = 1,\ldots,T ~\text{and}~h = h_{t-1}+1, \ldots, h_t: \nonumber \\
&                  & b^+_h(\bm{\omega}_{1:t}) + b^-_h(\bm{\omega}_{1:t}) \leq P, \nonumber \\
&                  & e_h(\bm{\omega}_{1:t}) = e_{h-1}(\bm{\omega}_{1:t}) + b^+_h(\bm{\omega}_{1:t}) \cdot \eta^+ \nonumber \\
&                  & ~~~~~~~~~~~~ - b^-_h(\bm{\omega}_{1:t}) / \eta^-, \nonumber \\
&                  & 0 \leq e_h(\bm{\omega}_{1:t}) \leq E, \nonumber \\
&\text{variables:} & \text{under each}~\bm{\omega}_{2:T} \in \{0,1\}^{T-1}, \nonumber \\ &                  & \quad\bm{b}^+_{t}(\bm{\omega}_{1:t})~\text{and}~\bm{b}^-_{t}(\bm{\omega}_{1:t}),~t=1,\ldots,T. 
\end{eqnarray}
Note that the expectation is taken over the random event schedules starting from day 2, namely $\bm{\omega}_{2:T}$, instead of $\bm{\omega}_{1:T}$. In practice, a customer is notified whether the next day is an event day at least a few hours ahead. Therefore, at the time of solving the problem, we already know the realization of $\omega_1$.

The problem in \eqref{eqn:optimization_problem} is a multi-stage stochastic decision problem. The numbers of decision variables and constraints are large and grow exponentially with the length of the time horizon. More specifically, for each day $t\in\{1,\ldots,T\}$, there are $2^{t-1}$ different scenarios (i.e., different realizations of $\bm{\omega}_{2:t}$), and therefore $2^{t-1}$ different battery schedules $\bm{b}^+_{t}(\bm{\omega}_{1:t})$ and $\bm{b}^-_{t}(\bm{\omega}_{1:t})$. So the total number of decision variables is 
$$
2 \cdot 24 \cdot \left( 1 + 2 + \cdots + 2^{T-1} \right) = 48 \cdot \left( 2^{T}-1 \right).
$$
For a one-year time horizon (i.e., $T=365$), there are over $10^{111}$ decision variables. The number of constraints grow exponentially with the time horizon in the same way.
In summary, it is impossible to obtain the exact solution to \eqref{eqn:optimization_problem} under any reasonable length of time horizon. In the next section, we show how to obtain a near-optimal solution.

\section{Proposed Solution}\label{sec:solution}
Since it is impractical to obtain the exact solution to \eqref{eqn:optimization_problem}, we get near-optimal solutions by a multistage model predictive control approach with a receding horizon of $N$ days, where $N \leq T$. Specifically, for each day $t$, we solve a stochastic program that minimizes the customer's total cost from day $t$ to day $t+N-1$, and obtain the battery schedules $\left\{\bm{b}^+_{\tau}, \bm{b}^-_{\tau} \right\}_{\tau=t}^{t+N-1}$. We use the decisions $\bm{b}^+_{t}, \bm{b}^-_{t}$ for the day $t$ only. For the next day $t+1$, we solve a stochastic program that minimizes the cost from day $t+1$ to day $t+N$, in order to get the schedules $\bm{b}^+_{t+1}, \bm{b}^-_{t+1}$. We repeat this process for each day. The hypothesis behind this approach is that when the receding horizon $N$ is sufficiently large, the current day's action is less likely to affect the costs more than $N$ days in the future. Therefore, the battery schedules obtained from the stochastic program over the smaller time horizon of $N$ days are likely to be close to the optimal schedules. Since the receding horizon (e.g., $N=30$ days) can be much smaller than the entire time horizon (e.g., $T=365$ days), we solve a much smaller problem than \eqref{eqn:optimization_problem}.

In addition to using a receding horizon of $N$ days, we further simplify the stochastic program by approximating the objective function in \eqref{eqn:optimization_problem} through sampling. The objective function in \eqref{eqn:optimization_problem} is the expected net cost, where the expectation is taken over the random event schedules $\bm{\omega}_{2:T}$. Such an expectation is computed as the weighted sum of net costs in $2^{T-1}$ different {\it scenarios}, where each scenario is a realization $\bm{\omega}_{1:T}$. In the multistage model predictive control approach outlined above, we limit the lookahead horizon to $N$ days, and take the expectation over the realization $\bm{\omega}_{t+1:t+N-1}$ (we already know $\omega_{t}$ when solving the problem). Therefore, we need to evaluate $2^{N-1}$ scenarios. Although we have $N < T$, the number of scenarios may still be large. This motivates us to further reduce the number of scenarios to consider.

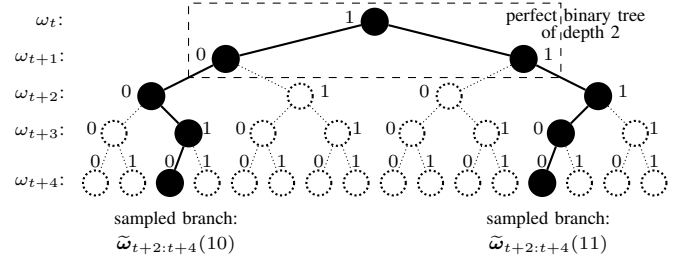
\begin{figure}
  \centering
  \begin{tikzpicture}[
   scale=0.33,
   root/.style={
      circle,
      minimum size=1mm,
      very thick,
      draw=black, 
      fill=black, 
      font=\scriptsize
      },
   node/.style={
      circle,
      minimum size=1mm,
      thick,
      densely dotted,
      draw=black, 
      fill=white, 
      font=\scriptsize
      },
   branch1/.style={
      circle,
      minimum size=6mm,
      very thick,
      draw=red!50, 
      fill=red!20, 
      font=\large
      },
   branch2/.style={
      circle,
      minimum size=6mm,
      very thick,
      draw=blue!50, 
      fill=blue!20, 
      font=\large
      },
   branch3/.style={
      circle,
      minimum size=6mm,
      very thick,
      draw=black!50, 
      fill=black!20, 
      font=\large
      },
   output/.style={
      rectangle,minimum size=6mm,rounded corners=3mm,
     very thick,draw=black,
     font=\large},
]

   \node (node1) at (0, 0) [root] {};
   \node (node2-1) at (-6, -1.5) [root] {};
   \node (node2-2) at (6, -1.5) [root] {};
   \node (node3-1) at (-9, -3) [root] {};
   \node (node3-2) at (-3, -3) [node] {};
   \node (node3-3) at (3, -3) [node] {};
   \node (node3-4) at (9, -3) [root] {};
   \node (node4-1) at (-10.5, -4.5) [node] {};
   \node (node4-2) at (-7.5, -4.5) [root] {};
   \node (node4-3) at (-4.5, -4.5) [node] {};
   \node (node4-4) at (-1.5, -4.5) [node] {};
   \node (node4-5) at (1.5, -4.5) [node] {};
   \node (node4-6) at (4.5, -4.5) [node] {};
   \node (node4-7) at (7.5, -4.5) [root] {};
   \node (node4-8) at (10.5, -4.5) [node] {};
   \node (node5-1) at (-11.25, -6.5) [node] {};
   \node (node5-2) at (-9.75, -6.5) [node] {};
   \node (node5-3) at (-8.25, -6.5) [root] {};
   \node (node5-4) at (-6.75, -6.5) [node] {};
   \node (node5-5) at (-5.25, -6.5) [node] {};
   \node (node5-6) at (-3.75, -6.5) [node] {};
   \node (node5-7) at (-2.25, -6.5) [node] {};
   \node (node5-8) at (-0.75, -6.5) [node] {};
   \node (node5-9) at (0.75, -6.5) [node] {};
   \node (node5-10) at (2.25, -6.5) [node] {};
   \node (node5-11) at (3.75, -6.5) [node] {};
   \node (node5-12) at (5.25, -6.5) [node] {};
   \node (node5-13) at (6.75, -6.5) [root] {};
   \node (node5-14) at (8.25, -6.5) [node] {};
   \node (node5-15) at (9.75, -6.5) [node] {};
   \node (node5-16) at (11.25, -6.5) [node] {};

   \draw [thick] (node1) -- (node2-1);
   \draw [thick] (node1) -- (node2-2);
   \draw [thick] (node2-1) -- (node3-1);
   \draw [densely dotted] (node2-1) -- (node3-2);
   \draw [densely dotted] (node2-2) -- (node3-3);
   \draw [thick] (node2-2) -- (node3-4);
   \draw [densely dotted] (node3-1) -- (node4-1);
   \draw [thick] (node3-1) -- (node4-2);
   \draw [densely dotted] (node3-2) -- (node4-3);
   \draw [densely dotted] (node3-2) -- (node4-4);
   \draw [densely dotted] (node3-3) -- (node4-5);
   \draw [densely dotted] (node3-3) -- (node4-6);
   \draw [thick] (node3-4) -- (node4-7);
   \draw [densely dotted] (node3-4) -- (node4-8);
   \draw [densely dotted] (node4-1) -- (node5-1);
   \draw [densely dotted] (node4-1) -- (node5-2);
   \draw [thick] (node4-2) -- (node5-3);
   \draw [densely dotted] (node4-2) -- (node5-4);
   \draw [densely dotted] (node4-3) -- (node5-5);
   \draw [densely dotted] (node4-3) -- (node5-6);
   \draw [densely dotted] (node4-4) -- (node5-7);
   \draw [densely dotted] (node4-4) -- (node5-8);
   \draw [densely dotted] (node4-5) -- (node5-9);
   \draw [densely dotted] (node4-5) -- (node5-10);
   \draw [densely dotted] (node4-6) -- (node5-11);
   \draw [densely dotted] (node4-6) -- (node5-12);
   \draw [thick] (node4-7) -- (node5-13);
   \draw [densely dotted] (node4-7) -- (node5-14);
   \draw [densely dotted] (node4-8) -- (node5-15);
   \draw [densely dotted] (node4-8) -- (node5-16);
   
   \node (day1) at (-13.1, 0) [] {\scriptsize $\omega_t$:};
   \node (day2) at (-13.5, -1.5) [] {\scriptsize $\omega_{t+1}$:};
   \node (day3) at (-13.5, -3) [] {\scriptsize $\omega_{t+2}$:};
   \node (day4) at (-13.5, -4.5) [] {\scriptsize $\omega_{t+3}$:};
   \node (day5) at (-13.5, -6.5) [] {\scriptsize $\omega_{t+4}$:};
   \node (scenario1) at (-1, 0.2) [] {\scriptsize $1$};
   \node (scenario2-1) at (-7, -1.3) [] {\scriptsize $0$};
   \node (scenario2-2) at (7, -1.3) [] {\scriptsize $1$};
   \node (scenario3-1) at (-10, -2.8) [] {\scriptsize $0$};
   \node (scenario3-2) at (-2, -2.8) [] {\scriptsize $1$};
   \node (scenario3-3) at (2, -2.8) [] {\scriptsize $0$};
   \node (scenario3-4) at (10, -2.8) [] {\scriptsize $1$};
   \node (scenario4-1) at (-11.4, -4.3) [] {\scriptsize $0$};
   \node (scenario4-2) at (-6.75, -4.3) [] {\scriptsize $1$};
   \node (scenario4-3) at (-5.4, -4.3) [] {\scriptsize $0$};
   \node (scenario4-4) at (-0.7, -4.3) [] {\scriptsize $1$};
   \node (scenario4-5) at (0.65, -4.3) [] {\scriptsize $0$};
   \node (scenario4-6) at (5.25, -4.3) [] {\scriptsize $1$};
   \node (scenario4-7) at (6.6, -4.3) [] {\scriptsize $0$};
   \node (scenario4-8) at (11.25, -4.3) [] {\scriptsize $1$};
   \node (scenario5-1) at (-11.25, -5.6) [] {\scriptsize $0$};
   \node (scenario5-2) at (-9.75, -5.6) [] {\scriptsize $1$};
   \node (scenario5-3) at (-8.25, -5.6) [] {\scriptsize $0$};
   \node (scenario5-4) at (-6.75, -5.6) [] {\scriptsize $1$};
   \node (scenario5-5) at (-5.25, -5.6) [] {\scriptsize $0$};
   \node (scenario5-6) at (-3.75, -5.6) [] {\scriptsize $1$};
   \node (scenario5-7) at (-2.25, -5.6) [] {\scriptsize $0$};
   \node (scenario5-8) at (-0.75, -5.6) [] {\scriptsize $1$};
   \node (scenario5-1) at (11.25, -5.6) [] {\scriptsize $1$};
   \node (scenario5-2) at (9.75, -5.6) [] {\scriptsize $0$};
   \node (scenario5-3) at (8.25, -5.6) [] {\scriptsize $1$};
   \node (scenario5-4) at (6.75, -5.6) [] {\scriptsize $0$};
   \node (scenario5-5) at (5.25, -5.6) [] {\scriptsize $1$};
   \node (scenario5-6) at (3.75, -5.6) [] {\scriptsize $0$};
   \node (scenario5-7) at (2.25, -5.6) [] {\scriptsize $1$};
   \node (scenario5-8) at (0.75, -5.6) [] {\scriptsize $0$};
   \draw [draw=black, dashed] (-7.5,-2.25) rectangle (7.5,0.75);
   \node at (8.1, 0.2) [] {\scriptsize perfect binary tree};
   \node at (8.3, -0.5) [] {\scriptsize of depth 2};
   \node at (-8, -8) [] {\scriptsize sampled branch:};
   \node at (-8, -9) [] {\scriptsize $\widetilde{\bm{\omega}}_{t+2:t+4}(10)$};
   \node at (7, -8) [] {\scriptsize sampled branch:};
   \node at (7, -9) [] {\scriptsize $\widetilde{\bm{\omega}}_{t+2:t+4}(11)$};

  \end{tikzpicture}

\caption{An example of the binary scenario tree. Day $t$ is an event day, the receding horizon is $N=5$, all possible scenarios $\bm{\omega}_{t:t+1}$ in the first $n=2$ days are evaluated, and one scenario $\widetilde{\bm{\omega}}_{t+2:t+4}(\bm{\omega}_{t:t+1})$ from day $t+2$ to day $t+4$ is sampled after each $\bm{\omega}_{t:t+1}$.}
\label{fig:example-networks}

\end{figure}

To better explain our approach, we can define a {\it binary scenario tree} to represent the scenarios to evaluate. The root indicates whether day $t$ is an event day, and the two child nodes of the root represent the two possibilities that day $t+1$ is an event day or a non-event day. In general, there are $2^{\tau-1}$ nodes of depth $\tau \geq 2$, representing the possibilities of day $t+\tau-1$ following the $2^{\tau-2}$ scenarios $\bm{\omega}_{t:t+\tau-2}$ from day $t$ to day $t+\tau-2$. We will have a {\it perfect binary tree} of depth $n \leq N$, in which all interior nodes have two children and all leaves have the same depth. Then each leaf node of this $n$-level perfect binary tree has one {\it degenerate branch} of depth $N-n$, namely a branch in which all interior nodes have exactly one child. Each degenerate branch corresponds to the selection of one scenario out of $2^{N-n}$ scenarios $\bm{\omega}_{t+n:t+N-1}$ from day $t+n$ to day $t+N-1$. We select the scenario by sampling according to the probabilities that each scenario occurs, where the probability of the scenario $\bm{\omega}_{t+n:t+N-1}$ is
\begin{eqnarray}
\prod_{\tau = t+n}^{t+N-1} (1-P_\tau)^{1-\omega_\tau} (P_\tau)^{\omega_\tau}
\end{eqnarray}

To define the reformulated problem to solve, for each $\tau \geq t+n$, we write $\widetilde{\bm{\omega}}_{t+n:\tau}(\bm{\omega}_{t:t+n-1})$ as the sampled realization of event days from day $t+n$ to day $\tau$ given the realization $\bm{\omega}_{t:t+n-1}$ of event days from day $t$ to day $t+n-1$. We write it as function of $\bm{\omega}_{t:t+n-1}$, because we need to sample one scenario after each $\bm{\omega}_{t:t+n-1}$. In the scenario tree, the selected $\widetilde{\bm{\omega}}_{t+n:t+N-1}(\bm{\omega}_{t:t+n-1})$ is the degenerate branch spawning from the leaf node of the branch $\bm{\omega}_{t:t+n-1}$ in the perfect binary tree of depth $n$. Note that $\widetilde{\bm{\omega}}_{t+n:t+N-1}(\bm{\omega}_{t:t+n-1})$ is random. For each $\tau \geq t+n$, we write $\widetilde{\bm{\omega}}_{t:\tau}(\bm{\omega}_{t:t+n-1})$ as the concatenation of $\bm{\omega}_{t:t+n-1}$ and $\widetilde{\bm{\omega}}_{t+n:\tau}(\bm{\omega}_{t:t+n-1})$.

Based on the above notations, we can write the decision variables in the multistage MPC problem as
\begin{eqnarray}\label{eqn:decision_variables_MPC}
&&\text{under each}~\bm{\omega}_{t+1:t+n-1} \in \{0,1\}^{n-1}, \\
&&\quad \text{for}~\tau=t,\ldots,t+n-1: \nonumber\\
&&\qquad \bm{b}^+_{\tau}(\bm{\omega}_{t:\tau}), \bm{b}^-_{\tau}(\bm{\omega}_{t:\tau}), \nonumber \\
&&\quad \text{for}~\tau= t+n, \ldots, t+N-1: \nonumber \\
&&\qquad\bm{b}^+_{\tau}\left[\widetilde{\bm{\omega}}_{t:\tau}(\bm{\omega}_{t:t+n-1})\right], 
                     \bm{b}^-_{\tau}\left[\widetilde{\bm{\omega}}_{t:\tau}(\bm{\omega}_{t:t+n-1})\right]. \nonumber
\end{eqnarray}
In other words, we need to determine the battery schedules under all possible scenarios from day $t+1$ to day $t+n-1$, and the schedules under only sampled scenarios from day $t+n$ to day $t+N-1$. Hence, we limit the number of variables to
\begin{eqnarray}
& & 2 \cdot 24 \cdot \left[ 1 + 2 + \cdots + 2^{n-1} + 2^{n-1} \cdot (N-n) \right] \nonumber \\
&=& 48 \cdot \left[ 2^{n-1} \cdot (N-n+2) - 1 \right].
\end{eqnarray}

We can see that the number of decision variables does not depend on the full length $T$ of the time horizon, and grows linearly with the receding horizon $N$ and exponentially with $n$. We control the complexity of the MPC problem through $n$.

Next, we define the modified net cost over the receding horizon. We write $\mathcal{I}_t$ as the set of intervals included in the receding time horizon from day $t$ to day $t+N-1$. For example, if $N=20$ and each interval is a month, then $\mathcal{I}_1 = \{1\}$ and $\mathcal{I}_{15} = \{1,2\}$. Dropping the dependence of the decision variables on the scenarios for notational simplicity, we can write the modified net cost as
\begin{eqnarray}\label{eqn:approximate_net_cost}
& & \widetilde{C}_t\left( \left\{\bm{b}^+_{\tau}, \bm{b}^-_{\tau} \right\}_{\tau=t}^{t+N-1} ; \bm{\omega}_{t:t+n-1} \right) \\
&=& \sum_{h=h_{t-1}+1}^{h_{t+N-1}} (c_h - p^e_h) \nonumber \\
&-& \sum_{i \in \mathcal{I}_t} \tilde{p}^{DRc}_i\left(\bm{b}^+_{t:t+N-1},\bm{b}^-_{t:t+N-1};\bm{\omega}_{t:t+n-1}\right) \nonumber \\
&-& \sum_{i \in \mathcal{I}_t} \sum_{\tau \in \mathcal{T}_i^E \cap [t,t+N-1]} p^{DRe}_\tau\left(\bm{b}^+_{t:\tau},\bm{b}^-_{t:\tau};\bm{\omega}_{t:t+n-1}\right), \nonumber
\end{eqnarray}
where the energy purchase cost $c_h$, the energy export payment $p_h^e$, and the DR energy payment $p^{DRe}_\tau$ remain the same as in \eqref{eqn:net_cost}, but the DR capacity payment $\tilde{p}^{DRc}_i$ is the approximation of the true DR capacity payment $p^{DRc}_i$. In our MPC approach, we may not be able to compute the true DR capacity payment, because the receding horizon may end before the end of an interval. Since we do not know the demand reduction in the remaining days of this interval, we cannot compute the average demand reduction in this interval and the true DR capacity payment. In this case, we calculate the DR capacity payment $p_i^{DRc}$ of interval $i$ according to \eqref{eqn:pdrkw} over interval $i$'s days before the end of the horizon, and then discount it by the fraction of total days in the interval that are considered, namely
\begin{eqnarray}
\tilde{p}^{DRc}_i = \frac{\min\{t_i,t+N-1\} - t_{i-1}}{t_i - t_{i-1}} \cdot p^{DRc}_i.
\end{eqnarray}
We discount the DR capacity payment, because the tariff costs and DR energy payments are calculated over a fraction of the month, and we want to use a corresponding portion of the total DR capacity payment over the month. If the receding horizon starts after the start of an interval, we use stored values of demand reductions in event days prior to the current receding horizon.

Finally, it is worth to note that the modified cost $\widetilde{C}_t\left( \left\{\bm{b}^+_{\tau}, \bm{b}^-_{\tau} \right\}_{\tau=t}^{t+N-1} ; \bm{\omega}_{t:t+n-1} \right)$ is random, because we randomly select the scenario $\widetilde{\bm{\omega}}_{t+n:t+N-1}(\bm{\omega}_{t:t+n-1})$ after day $t+n-1$.

Now we can formulate the problem to solve in our MPC approach as follows.
\begin{eqnarray} \label{eqn:approximate_optimization_problem}
&\text{min}  &
\!\!\!\!\!\!\!\!\mathbb{E}_{\bm{\omega}_{t+1:t+n-1}} \left\{
\widetilde{C}_t\left( \left\{\bm{b}^+_{\tau}, \bm{b}^-_{\tau} \right\}_{\tau=t}^{t+N-1} ; \bm{\omega}_{t:t+n-1} \right) 
\right\} \nonumber \\
&s. t. & \!\!\!\!\!\!\!\!\text{under each}~\bm{\omega}_{t+1:t+n-1} \in \{0,1\}^{n-1}, \nonumber \\
&                  & \!\!\!\!\text{for}~\tau = t,\ldots,t+n-1,~\text{and} \nonumber \\
&                  & ~~~h = h_{\tau-1}+1, \ldots, h_\tau: \nonumber \\
&                  & b^+_h(\bm{\omega}_{t:\tau}) + b^-_h(\bm{\omega}_{t:\tau}) \leq P, \nonumber \\
&                  & e_h(\bm{\omega}_{t:\tau}) = e_{h-1}(\bm{\omega}_{t:\tau}) + b^+_h(\bm{\omega}_{t:\tau}) \cdot \eta^+ \nonumber \\
&                  & ~~~~~~~~~~~~~~~ - b^-_h(\bm{\omega}_{t:\tau}) / \eta^-, \nonumber \\
&                  & 0 \leq e_h(\bm{\omega}_{t:\tau}) \leq E, \nonumber \\
&                  & \!\!\!\!\text{for}~\tau = t+n,\ldots,t+N-1 ~\text{and} \nonumber\\
&                  & ~~~h = h_{\tau-1}+1, \ldots, h_\tau: \nonumber \\
&                  & \!b^+_h\left[\widetilde{\bm{\omega}}_{t:\tau}(\bm{\omega}_{t:t+n-1})\right] +
                        b^-_h\left[\widetilde{\bm{\omega}}_{t:\tau}(\bm{\omega}_{t:t+n-1})\right] \leq P, \nonumber \\
&                  & \!e_h\left[\widetilde{\bm{\omega}}_{t:\tau}(\bm{\omega}_{t:t+n-1})\right] =
                        e_{h-1}\left[\widetilde{\bm{\omega}}_{t:\tau}(\bm{\omega}_{t:t+n-1})\right] \nonumber \\
&                  & ~~~~~~~~~~~~~~~~~~~~~~ +b^+_h\left[\widetilde{\bm{\omega}}_{t:\tau}(\bm{\omega}_{t:t+n-1})\right] \cdot \eta^+ \nonumber \\
&                  & ~~~~~~~~~~~~~~~~~~~~~~ - b^-_h\left[\widetilde{\bm{\omega}}_{t:\tau}(\bm{\omega}_{t:t+n-1})\right] / \eta^-, \nonumber \\
&                  & \!0 \leq e_h\left[\widetilde{\bm{\omega}}_{t:\tau}(\bm{\omega}_{t:t+n-1})\right] \leq E, \nonumber \\
&\text{variables:} & \text{as defined in}~\eqref{eqn:decision_variables_MPC}.
\end{eqnarray}
Similar to \eqref{eqn:optimization_problem}, we take the expectation over the random event schedules starting from day $t+1$, namely $\bm{\omega}_{t+1:t+n-1}$, because we already know the realization of $\omega_t$ when solving the problem for day $t$.

\begin{remark}[Approximation and Randomness]\label{remark:solution}
Note that the objective function and decision variables in the proposed MPC approach \eqref{eqn:approximate_optimization_problem} are different from those in the original scheduling problem \eqref{eqn:optimization_problem}. For computational feasibility, we reduce the number of decision variables by reducing the time horizon and evaluating only selected scenarios, and make necessary approximations in the cost function due to the reduced time horizon. Therefore, the proposed solution \eqref{eqn:approximate_optimization_problem} is an approximation to the optimal solution to \eqref{eqn:optimization_problem}. We can recover the optimal solution by setting the time horizon the same as that in the original problem (i.e., $N=T$) and minimizing the exact expected cost (i.e., $n=N$).
Note also that the objective function and the constraints in \eqref{eqn:approximate_optimization_problem} are random due to the randomly sampled $\widetilde{\bm{\omega}}_{t+n:t+N-1}(\bm{\omega}_{t:t+n-1})$ after each $\bm{\omega}_{t:t+n-1}$ (even though we take the expectation over $\bm{\omega}_{t+1:t+n-1}$). Therefore, the solution, namely the battery schedules, is random and depends on which scenario $\widetilde{\bm{\omega}}_{t+n:t+N-1}(\bm{\omega}_{t:t+n-1})$ was sampled. In Section V-B, we study the effect of random sampling numerically and find it to be negligible for the cases we investigated. We will also study the suboptimality introduced by limiting the time horizon to $N$ in the next section.
\end{remark}

\section{Case Study}\label{CASE STUDY}
In this section, we apply the model and the solution in Section \ref{MODEL} and Section \ref{sec:solution} to an example residential customer participating in a tariff and DR program based on the existing Customer Grid Supply Plus tariff \cite{HawaiianElectric} and the Capacity Reduction Grid Service DR program \cite{HawaiianElectric2018}\footnote{This case study simplifies some elements of the Customer Grid Supply Plus tariff and the Capacity Reduction grid service, but aims to represent the major elements that could lead to incentives to game baselines. For example, Capacity Reduction includes separate payments to aggregators and customers, but this study considers that the aggregator and customers act as one entity. Interested readers can refer to the program documents for more information.}. 

We first study several cases whose time horizons are small enough for us to compute the optimal solutions, and compare our proposed solutions with the optimal solutions. Next, we present numerical results for a case study with a full-year time horizon, in order to identify customer incentives to increase baseline energy consumption.

\subsection{Basic Simulation Setup}
We first describe the parameters that are common to all of the cases. 

\subsubsection{Customer default electricity demand} The default customer electricity demand is the base-case Honolulu residential customer demand obtained from an OpenEI dataset \cite{OfficeofEnergyEfficiency&RenewableEnergyEERE}, which includes synthesized hourly load profiles for a typical year's weather corresponding to the Typical Meteorological Year 3 (TMY3) dataset \cite{NationalRenewableEnergyLaboratorya}. 

\subsubsection{Solar photovoltaic energy production} The solar production is obtained from PVWatts \cite{NationalRenewableEnergyLaboratory}, based on a 9.5 kW rooftop system with default parameters and the Honolulu TMY3 weather data. Using TMY3 data for both customer demand and solar aligns their weather-based variations. The solar system size was selected so that annual production roughly matches annual customer demand. 

\subsubsection{Battery energy storage system} The customer's battery consists of two units of the Tesla Powerwall battery \cite{Tesla}, a popular residential battery system. The round-trip efficiency is 90\%. Assuming equal charging and discharging efficiency, we have $\eta^+ = \eta^- = \sqrt{0.9}$. The total battery system power capacity is 10 kW and energy capacity is 27 kWh. The battery energy capacity was selected to be similar to the customer's daily energy demand.

\subsubsection{Customer tariff} The customer receives \$0.108/kWh for energy exported to the grid, and the customer pays \$0.29/kWh to purchase energy from the grid. 
The difference between the export rate and the purchase rate incentivizes charging the battery when solar production exceeds load, and discharging that energy later to reduce grid purchases.

\subsubsection{DR program details} The customer receives a DR capacity payment based on average load reduction during DR events each month. The DR window is from 5 p.m. to 9 p.m. in all event days. 
Baseline energy consumption is the average energy consumption during the DR windows of some number of previous non-event days. 

\subsubsection{Probabilities of DR events} The probabilities of a DR event occurring each day were calculated based on a logistic regression of occurrence of a DR event versus daily peak temperature. The logistic regression model was created based on 2017 Hawaiian Electric system electricity demand \cite{FederalEnergyRegulatoryCommission} and 2017 weather data for Honolulu \cite{WeatherUnderground}. Capacity Reduction DR events can be called up to 104 times per year \cite{HawaiianElectric2018}, so event days were assigned to the 104 days of 2017 with the highest peak hourly electricity demand. A logistic regression model of DR event occurrence versus daily peak temperature was used to calculate daily event probabilities for Honolulu TMY3 weather. 

\subsubsection{Initial values} For the first day, the initial battery state of charge is set to 50\% of the battery energy capacity, and the initial baseline is set to all zeros. 

\subsection{Assessing The Quality of Proposed Solutions}
As explained in Remark~\ref{remark:solution}, our proposed solution is an approximation to the optimal solution, and is random. Therefore, it is important to assess its performance in terms of the cost and its variance. To compare with the optimal solution, we need to restrict to cases whose time horizons are small enough so that it is feasible to compute the optimal solution.

We consider a time horizon of $T=7$ days, which is long enough to show multi-day effects, and short enough to allow computing the optimal solution. 
Due to the small number of days in these cases, we use a 3-day baseline (i.e., the baseline consumption is the average of previous 3 non-event days), which is shorter than the standard 10-day baseline in Capacity Reduction. We consider two periods of 7 days: the first 7 days in January (low temperature and low probabilities of events) and the first 7 days in October (high temperature and high probabilities of events). We set the rate $r^{DRc}_i$ of DR capacity payments to be \$2/kW (the minimum customer incentive level allowed for Capacity Reduction grid service \cite{HawaiianElectric2018}), and use the per-hour demand reduction averaged over the 7 days, instead of over the entire month, to determine DR capacity payments.

Table~\ref{table:1} presents the results for the 7-day case study in January and October. The table shows customer cost and demand response quantity, as expected values over the random event schedules $\bm{\omega}_{1:7}$. Additional metrics baseline load (i.e., average baseline load over all events in kW) and event load (i.e., average load over all events in kW) were also calculated. 

We first show the metrics under the optimal solution, which is obtained by setting $N=n=7$. Then we evaluate our proposed solution under different receding horizons of $N=4$ and $N=2$ without sampling (i.e., $n=N$ in  both cases). In these two cases with $n=N$, we optimize the expected cost over all event schedules $\bm{\omega}_{t:t+N-1}$, which results in deterministic solutions. In this way, we can focus on the impact of a shorter receding horizon. We can see that under the 4-day receding horizon, customer cost nearly matches the optimal cost, and under the 2-day receding horizon, customer cost is about 3\% higher than the optimal cost. This suggests that a shorter receding horizon can provide close to optimal results. As we would expect, the performance tends to improve with the length of the receding horizon. 

Next we focus on the impact of sampling the scenario tree. We set the receding horizon as the entire time horizon (i.e., $N=T=7$), consider all the possible scenarios in the first 2 days (i.e., $n=2$), and sample one event schedule from day $3$ to day $7$. In this case, the proposed solution depends on the sampled event schedule $\widetilde{\bm{\omega}}_{3:7}$, and therefore, is random. We run the simulations 5 times to get 5 battery schedules under different sampled event schedules, and show the mean and the standard deviation of the performance metrics over these 5 runs. We can see that the customer cost is within 1\% of the optimum with small standard deviation.

Finally, we set $N=4$ and $n=2$ to evaluate the proposed solution with both a shorter receding horizon and sampled event schedules. Again, since the proposed approach yields different battery schedules under different sampled event schedules, we show the average metrics and standard deviations over 5 runs. The customer cost is within 1\% of the optimum with small standard deviation.

Comparing the cases of $(N,n)=(7,2)$ and $(N,n)=(4,2)$ with the case of $(N,n)=(2,2)$, we can see that under the same depth of the perfect binary tree (i.e., $n=2$), we can improve the performance (i.e., from about 3\% within the optimum to within 1\% from the optimum) by sampling additional days (i.e., 5 additional days when $N=7$ and 2 additional days when $N=4$). 

Compared to customer cost, the DR quantity metric tends to show greater percentage differences from the optimal case. However, the values are still reasonably close to the optimal case in absolute magnitude and have reasonably small standard deviations. This finding also holds for the baseline load and event load metrics. 

Our conclusion from this study is that it is possible to significantly reduce the number of scenarios considered in the scenario tree and still obtain near-optimal results. We thus use our proposed approach in the full-year study, where calculating the optimal solution is not feasible. 

\begin{table}
  \caption{Approximation study results}
  \label{table:1}
  \includegraphics[width=\linewidth]{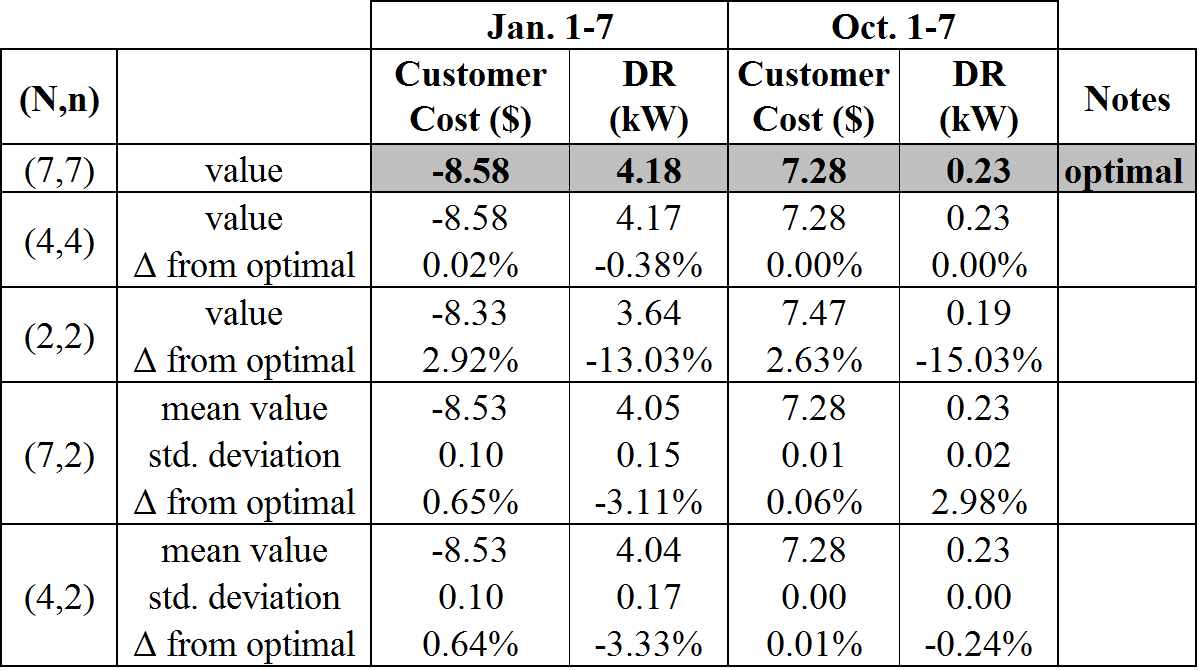}
\end{table}



\subsection{Evaluating Customer Incentives}
We study the battery scheduling problem with a time horizon of a full year (i.e., $T=365$). In the proposed solution, we choose a receding horizon of $N = 35$ (longer than a full month, over which the DR capacity payment is calculated) and a depth $n = 4$ of the perfect binary scenario tree (based on results in Table~\ref{table:1}). We consider two DR capacity payment rates of \$2/kW and \$10/kW. 

We also create a ``counter-factual" case, as a control group, 
where the customer does not participate in the DR program. In this case the objective is to minimize net tariff costs (i.e., costs of energy purchase from the grid minus rebates for exports). The stochastic parameters representing event schedule do not impact this cost, so the optimal decisions for this case can be solved in a single scenario. However, in the counter-factual case, there are many different solutions that yield the same optimal cost with varying levels of electricity consumption during DR windows. In order to compare event and baseline load between the DR cases and the counter-factual case, we need to select one from multiple optimal solutions. For this study, we selected the optimal battery dispatch schedule that charges the battery as much as possible to absorb solar energy that exceeds load, and discharges the battery as much as possible to avoid purchasing power from the grid. This schedule was selected because it is a realistic and simple algorithm to implement on batteries that are deployed today, and thus represents a plausible counter-factual case. 

Table~\ref{table:2} shows results for the cases with \$2/kW-month and \$10/kW-month DR capacity payments. In addition to the metrics shown, we also calculated average baseline load and event load. Since the proposed approach yields stochastic solutions that depend on the sampled schedules, we run the experiments 10 times to get the mean values of the metrics. For these cases we calculate the baseline inflation (the difference between baseline load with DR payments and baseline load in the case with no DR payment), and use this to calculate ``Baseline Inflation (\% of DR)," which shows the portion of apparent DR load reduction is actually due to baseline inflation. 

\begin{table}
  \caption{Full-year results}
  \label{table:2}
  \centering
  \includegraphics[width=\linewidth]{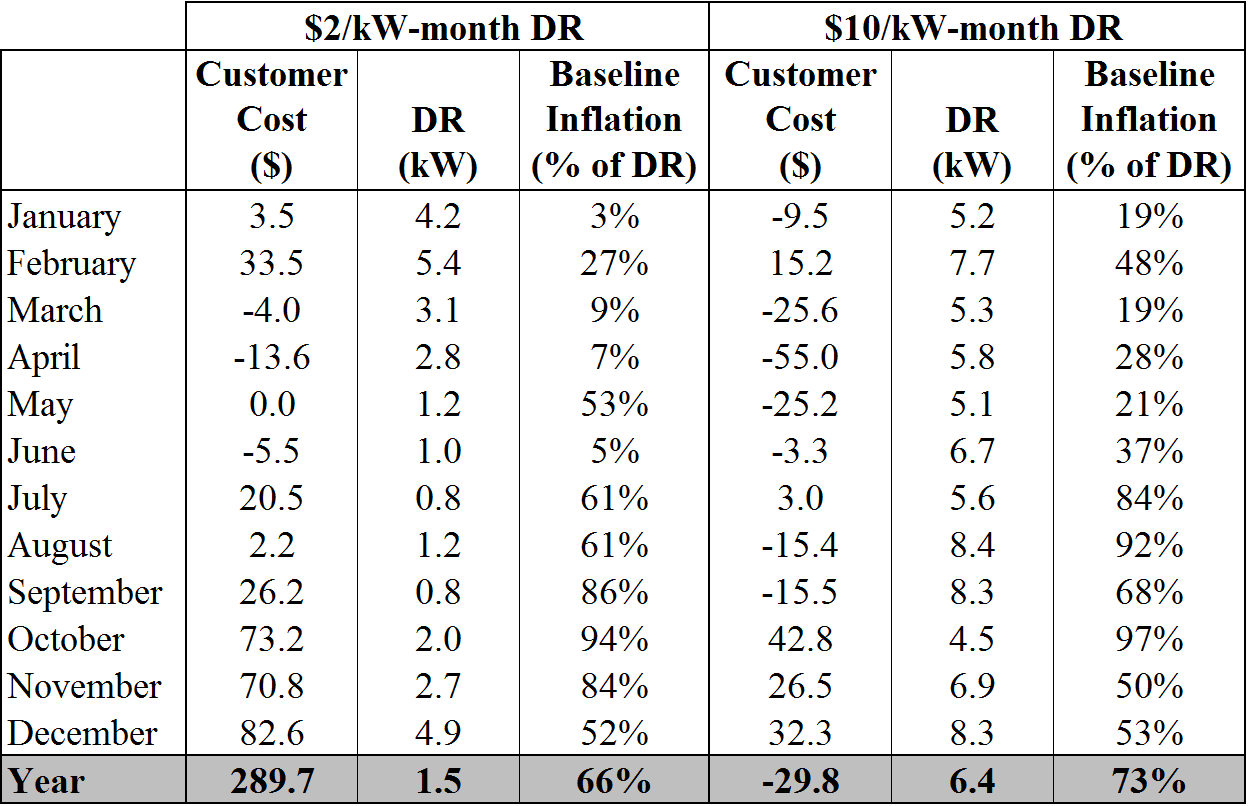}
\end{table}




In the ``counter-factual" case with no DR payments, average baseline load and event load for the year are both approximately 0.1 kW, and average DR quantity is approximately 0.0 kW. This reflects that when the battery is used to minimize tariff costs via our heuristic, the battery covers most of the load during the DR windows of 5 p.m. to 9 p.m.. There is not a significant difference in DR window energy consumption between event and non-event days. 

The \$2/kW-month case shows higher baseline load (1.1 kW) and lower event load (-0.4 kW). This indicates that while the DR program does incentivize the customer to reduce load during events, it also incentivizes even greater increases in energy consumption during DR windows of non-event days. Over the year, 66\% of the apparent DR is due to baseline inflation, with monthly values ranging from 3\% in January to 94\% in October. 

The \$2/kW-month case also indicates that even at the minimum payment level, the Capacity Reduction grid service successfully provides an incentive for the customer to modify battery charging to deliver DR. Comparing the quantity of average DR (1.5 kW) to the size of the battery system (10 kW, 27 kWh), we see that the quantity of DR is much smaller than what could be obtained if the battery was solely focused on delivering DR. This reflects that at this relatively low DR payment level, the battery often prioritizes the high-value application of shifting solar production to reduce tariff costs. 

The \$10/kW-month case shows even higher baseline load (4.8 kW) and lower event load (-1.7 kW), indicating stronger incentives to reduce load during events and increase load during DR windows of non-event days. Over the year, 73\% of the apparent DR is due to baseline inflation, with monthly values ranging from 19\% in January and March to 97\% in October. While the overall trend was to increase the percentage of baseline inflation compared to the \$2/kW-month case, in some months (May, September, November) the percentage of baseline inflation was lower. This is because the relationship between DR payment and strength of incentives to lower event load and increase baseline load can vary based on monthly parameters including event probabilities, customer default demand, and solar energy production.

\bibliographystyle{IEEEtran}
\bibliography{EE635_project.bib,Battery_DR_MPC.bib}

\begin{thebibliography}{10}
\providecommand{\url}[1]{#1}
\csname url@samestyle\endcsname
\providecommand{\newblock}{\relax}
\providecommand{\bibinfo}[2]{#2}
\providecommand{\BIBentrySTDinterwordspacing}{\spaceskip=0pt\relax}
\providecommand{\BIBentryALTinterwordstretchfactor}{4}
\providecommand{\BIBentryALTinterwordspacing}{\spaceskip=\fontdimen2\font plus
\BIBentryALTinterwordstretchfactor\fontdimen3\font minus
  \fontdimen4\font\relax}
\providecommand{\BIBforeignlanguage}[2]{{%
\expandafter\ifx\csname l@#1\endcsname\relax
\typeout{** WARNING: IEEEtran.bst: No hyphenation pattern has been}%
\typeout{** loaded for the language `#1'. Using the pattern for}%
\typeout{** the default language instead.}%
\else
\language=\csname l@#1\endcsname
\fi
#2}}
\providecommand{\BIBdecl}{\relax}
\BIBdecl

\bibitem{Vardakas2015}
J.~S. Vardakas, N.~Zorba, and C.~V. Verikoukis, ``{A Survey on Demand Response
  Programs in Smart Grids: Pricing Methods and Optimization Algorithms},''
  \emph{IEEE Communications Surveys and Tutorials}, vol.~17, no.~1, pp.
  152--178, 2015.

\bibitem{SmartElectricPoweralliiance2018}
\BIBentryALTinterwordspacing
B.~Chew, B.~Feldman, D.~Ghosh, and M.~Surampudy, ``{2018 Utility Demand
  Response Market Snapshot},'' Smart Electric Power Alliance, Tech. Rep., 2018.
  [Online]. Available:
  \url{https://sepapower.org/resource/2018-demand-response-market-snapshot/}
\BIBentrySTDinterwordspacing

\bibitem{Dobakhshari2018}
D.~G. Dobakhshari and V.~Gupta, ``{A Contract Design Approach for Phantom
  Demand Response},'' \emph{IEEE Transactions on Automatic Control}, p.~1,
  2018.

\bibitem{Muthirayan2019}
\BIBentryALTinterwordspacing
D.~Muthirayan, E.~Baeyens, P.~Chakraborty, K.~Poolla, and P.~P. Khargonekar,
  ``{A Minimal Incentive-based Demand Response Program With Self Reported
  Baseline Mechanism},'' 2019. [Online]. Available:
  \url{https://arxiv.org/pdf/1901.02923.pdf}
\BIBentrySTDinterwordspacing

\bibitem{muthirayan2017mechanism}
D.~Muthirayan, D.~Kalathil, K.~Poolla, and P.~Varaiya, ``{Mechanism design for
  demand response programs},'' \emph{arXiv preprint arXiv:1712.07742}, 2017.

\bibitem{Vuelvas2018}
J.~Vuelvas, F.~Ruiz, and G.~Gruosso, ``{Limiting gaming opportunities on
  incentive-based demand response programs},'' \emph{Applied Energy}, vol. 225,
  pp. 668--681, sep 2018.

\bibitem{Vuelvas2018a}
\BIBentryALTinterwordspacing
J.~Vuelvas and F.~Ruiz, ``{Rational consumer decisions in a peak time rebate
  program},'' feb 2018. [Online]. Available:
  \url{http://arxiv.org/abs/1802.08112}
\BIBentrySTDinterwordspacing

\bibitem{Ellman2019}
D.~Ellman and Y.~Xiao, ``{Customer Incentives for Gaming Demand Response
  Baselines},'' in \emph{Accepted by 58th IEEE Conference on Decision and
  Control}, 2019.

\bibitem{Bruno2018}
S.~Bruno, G.~Giannoccaro, and M.~L. Scala, ``{Optimization of residential
  storage and energy resources under demand response schemes},'' in \emph{2018
  19th IEEE Mediterranean Electrotechnical Conference (MELECON)}.\hskip 1em
  plus 0.5em minus 0.4em\relax IEEE, may 2018, pp. 225--230.

\bibitem{Garifi2018}
K.~Garifi, K.~Baker, B.~Touri, and D.~Christensen, ``{Stochastic Model
  Predictive Control for Demand Response in a Home Energy Management System},''
  in \emph{2018 IEEE Power {\&} Energy Society General Meeting (PESGM)}.\hskip
  1em plus 0.5em minus 0.4em\relax IEEE, aug 2018, pp. 1--5.

\bibitem{Castelo-Becerra2017}
A.~Castelo-Becerra, W.~Zeng, and M.-Y. Chow, ``{Cooperative distributed
  aggregation algorithm for demand response using distributed energy storage
  devices},'' in \emph{2017 North American Power Symposium (NAPS)}.\hskip 1em
  plus 0.5em minus 0.4em\relax IEEE, sep 2017, pp. 1--6.

\bibitem{Wolak2006}
\BIBentryALTinterwordspacing
F.~Wolak, ``{Residential Customer Response to Real-Time Pricing: The Anaheim
  Critical-Peak Pricing Experiment},'' 2006. [Online]. Available:
  \url{https://escholarship.org/uc/item/3td3n1x1}
\BIBentrySTDinterwordspacing

\bibitem{JimPierobon2013}
\BIBentryALTinterwordspacing
J.~Pierobon, ``{Two FERC settlements illustrate attempts to `game' demand
  response programs},'' 2013. [Online]. Available:
  \url{https://www.theenergyfix.com/2013/07/25/two-ferc-settlements-illustrate-attempts-to-game-demand-response-programs/}
\BIBentrySTDinterwordspacing

\bibitem{Vedullapalli2019}
D.~T. Vedullapalli, R.~Hadidi, and B.~Schroeder, ``{Optimal Demand Response in
  a building by Battery and HVAC scheduling using Model Predictive Control},''
  in \emph{2019 IEEE/IAS 55th Industrial and Commercial Power Systems Technical
  Conference (I{\&}CPS)}.\hskip 1em plus 0.5em minus 0.4em\relax IEEE, may
  2019, pp. 1--6.

\bibitem{Vrettos2013}
E.~Vrettos, K.~Lai, F.~Oldewurtel, and G.~Andersson, ``{Predictive Control of
  buildings for Demand Response with dynamic day-ahead and real-time prices},''
  in \emph{2013 European Control Conference (ECC)}.\hskip 1em plus 0.5em minus
  0.4em\relax IEEE, jul 2013, pp. 2527--2534.

\bibitem{Keerthisinghe2019}
C.~Keerthisinghe, A.~C. Chapman, and G.~Verbi{\v{c}}, ``{Energy Management of
  PV-Storage Systems: Policy Approximations Using Machine Learning},''
  \emph{IEEE Transactions on Industrial Informatics}, vol.~15, no.~1, pp.
  257--265, jan 2019.

\bibitem{Jin2017}
X.~Jin, K.~Baker, S.~Isley, and D.~Christensen, ``{User-preference-driven model
  predictive control of residential building loads and battery storage for
  demand response},'' in \emph{2017 American Control Conference (ACC)}.\hskip
  1em plus 0.5em minus 0.4em\relax IEEE, may 2017, pp. 4147--4152.

\bibitem{Xu2014}
Y.~Xu and L.~Tong, ``{On the operation and value of storage in consumer demand
  response},'' in \emph{53rd IEEE Conference on Decision and Control}.\hskip
  1em plus 0.5em minus 0.4em\relax IEEE, dec 2014, pp. 205--210.

\bibitem{HawaiianElectric2018}
\BIBentryALTinterwordspacing
{Hawaiian Electric}, ``{Addendum No. 2 To Request for Proposals For Provision
  of Grid Services Utilizing Demand-Side Resources},'' 2018. [Online].
  Available:
  \url{https://www.hawaiianelectric.com/documents/products{\_}and{\_}services/
  demand{\_}response/dr{\_}rfp{\_}best{\_}and{\_}final{\_}offer.pdf}
\BIBentrySTDinterwordspacing

\bibitem{ConEdison2019}
\BIBentryALTinterwordspacing
{Con Edison}, ``{Commercial Demand Response Program Guidelines},'' 2019.
  [Online]. Available:
  \url{https://www.coned.com/-/media/files/coned/documents/save-energy-money/rebates-incentives-tax-credits/smart-usage-rewards/smart-usage-program-guidelines.pdf?la=en}
\BIBentrySTDinterwordspacing

\bibitem{HawaiianElectric}
\BIBentryALTinterwordspacing
{Hawaiian Electric}, ``{Customer Grid-Supply Plus}.'' [Online]. Available:
  \url{https://www.hawaiianelectric.com/products-and-services/customer-renewable-programs/customer-grid-supply-plus}
\BIBentrySTDinterwordspacing

\bibitem{OfficeofEnergyEfficiency&RenewableEnergyEERE}
\BIBentryALTinterwordspacing
{Office of Energy Efficiency {\&} Renewable Energy (EERE)}, ``{Commercial and
  Residential Hourly Load Profiles for all TMY3 Locations in the United
  States}.'' [Online]. Available:
  \url{https://openei.org/doe-opendata/dataset/commercial-and-residential-hourly-load-profiles-for-all-tmy3-locations-in-the-united-states}
\BIBentrySTDinterwordspacing

\bibitem{NationalRenewableEnergyLaboratorya}
\BIBentryALTinterwordspacing
{National Renewable Energy Laboratory}, ``{National Solar Radiation Data Base:
  1991- 2005 Update: Typical Meteorological Year 3}.'' [Online]. Available:
  \url{https://rredc.nrel.gov/solar/old{\_}data/nsrdb/1991-2005/tmy3/}
\BIBentrySTDinterwordspacing

\bibitem{NationalRenewableEnergyLaboratory}
\BIBentryALTinterwordspacing
------, ``{PVWatts Calculator}.'' [Online]. Available:
  \url{https://pvwatts.nrel.gov/pvwatts.php}
\BIBentrySTDinterwordspacing

\bibitem{Tesla}
\BIBentryALTinterwordspacing
Tesla, ``{Powerwall | The Tesla Home Battery}.'' [Online]. Available:
  \url{https://www.tesla.com/powerwall}
\BIBentrySTDinterwordspacing

\bibitem{FederalEnergyRegulatoryCommission}
\BIBentryALTinterwordspacing
{Federal Energy Regulatory Commission}, ``{Form 714 - Annual Electric Balancing
  Authority Area and Planning Area Report}.'' [Online]. Available:
  \url{https://www.ferc.gov/docs-filing/forms/form-714/data.asp}
\BIBentrySTDinterwordspacing

\bibitem{WeatherUnderground}
\BIBentryALTinterwordspacing
{Weather Underground}, ``{Weather History {\&} Data Archive}.'' [Online].
  Available: \url{https://www.wunderground.com/history/}
\BIBentrySTDinterwordspacing

\end{thebibliography}


\end{document}